\documentstyle[twocolumn,aps,prl,epsfig]{revtex}

\draft

\begin{document}

\title{Diffractive Orbits in an Open Microwave Billiard}
\author{
  J. S. Hersch$^*$, M. R.  Haggerty$^*$, 
  and E. J. Heller$^\dagger$ \\}
\address{
  $^*$Department of Physics, Harvard University, Cambridge,
  Massachusetts 02138\\
  $^\dagger$Departments of Chemistry and Physics, 
  Harvard University, Cambridge, Massachusetts 02138}

\maketitle
\begin{abstract}
    We demonstrate the existence and significance of diffractive
    orbits in an open microwave billiard, both experimentally and
    theoretically.  Orbits that diffract off of a sharp edge strongly
    influence the conduction spectrum of this resonator, especially in
    the regime where there are no stable classical orbits.  On
    resonance, the wavefunctions are influenced by both classical and
    diffractive orbits.  Off resonance, the wavefunctions are
    determined by the constructive interference of multiple transient,
    nonperiodic orbits.  Experimental, numerical, and semiclassical
    results are presented.
\end{abstract}

\bigskip

\begin{figure}
    \centerline{\epsfig{figure=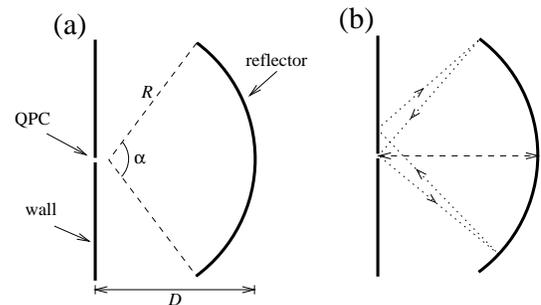,width=7cm}}
    \caption{
      (a) A schematic of the mesoscopic resonator studied by Katine,
      with radius of curvature $R$, opening angle $\alpha$, and
      reflector-wall separation $D$.  Electrons impinge on the wall
      from the left, and the conductance to the region on the right is
      measured.  (b) Two closed orbits of the unstable resonator:
      diffractive (dotted line), and horizontal (dashed line).  These
      will be discussed later in the paper.}
    \label{schematic}
\end{figure}

Recently, Katine et al. studied the transmission behavior of an open
quantum billiard in the context of a two dimensional electron gas
(2DEG) in a GaAs/AlGaAs heterostructure~\cite{katine}.  Their
resonator was formed by a wall with a small aperture, called a quantum
point contact (QPC), and an arc-shaped reflector.  A schematic of this
resonator is shown in Fig.~\ref{schematic}.  The voltage on the
reflector could be varied, effectively moving the reflector towards or
away from the wall.  Their measurements showed a series of conductance
peaks, analogous to those seen in a Fabry-Perot, as the reflector
position was varied.

As we discuss below, the resonator considered here represents a new
class of billiards, to our knowledge not previously studied in the
literature.  That is, the billiard is geometrically {\em open}, but in
the stable regime, it is classically {\em closed}.  In the unstable
regime, the resonance properties of the billiard are determined in
large part by {\em diffraction}.  

The resonator shown in Fig.~\ref{schematic} has two distinct modes of
operation.  When the center of curvature of the reflector is to the
left of the wall (the regime studied in~\cite{katine}), then all
classical paths starting from the QPC that hit the reflector remain
forever in the region between wall and the reflector: the dynamics is
stable and the periodic orbits can be semiclassically quantized.  Each
quantized mode of the resonator can be characterized by two quantum
numbers $(n,m)$, which represent the number of radial and angular
nodes respectively.  As the reflector-wall separation is varied, the
conductance exhibits a peak each time one of these quantized modes is
allowed.  Once an electron is in the resonator, the only way for it to
leave is by tunneling back through the QPC or by diffracting around
the reflector; since both processes are slow, the resonances have
narrow widths.  Because the QPC is on the symmetry axis, only modes
with even $m$ can be excited.

When the center of curvature is to the right of the wall, however, the
dynamics becomes unstable: all classical trajectories beginning at the
QPC rapidly bounce out of the resonator, except for a single unstable
periodic orbit along the axis of symmetry, which we will call the
``horizontal'' orbit [see Fig~\ref{schematic}(b)].  Although the
horizontal orbit returns to the QPC, it has a low probability of
escaping the resonator there because the QPC is much smaller than the
de Broglie wavelength of the electron.  Because the horizontal orbit
is the only periodic orbit in the unstable regime, one might expect
resonant buildup only along the symmetry axis.  Such a spectrum would
be quasi-one-dimensional, with only the half-wavelength periodicity of
a Fabry-Perot cavity.  However, in numerical simulations it was found
that there were other transmission resonances in the unstable regime
which did not correspond to any classical periodic
orbits~\cite{jonathan}.  It was proposed that these anomalous peaks
are supported by diffraction off the tips of the reflector.
Unfortunately, in the mesoscopic experiments, decoherence of the
electron wave by impurities in the GaAs/AlGaAs heterostructure
shortens the lifetime of the resonances, leaving insufficient energy
resolution to resolve the diffractive peaks~\cite{westervelt}.

\begin{figure}
    \centerline{\epsfig{figure=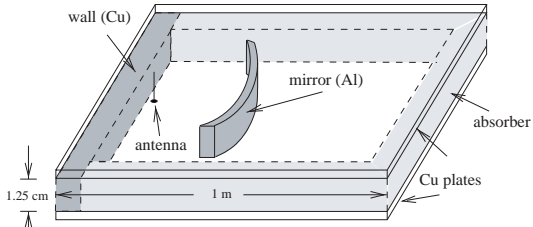, width=7cm}}
    \caption{
      This is the microwave analog of the mesoscopic resonator studied
      by Katine.  The antenna simulates the QPC; to reduce its
      coupling to the resonator, it is placed very close to the wall.
      The drawing is not to scale.  }
    \label{experiment}
\end{figure}

For this reason, we decided to investigate a parallel plate microwave
resonator with a similar geometry.  In microwave experiments,
decoherence and dissipation are not a problem, the geometry of the
resonator can be specified much more accurately, and the dynamical
range of available wavelengths is much larger.  The experimental setup
is shown in Fig.~\ref{experiment}.

For the transverse electromagnetic (TEM) mode, it can be shown that
the equation governing the component of the electric field normal to
the plates is identical to the two-dimensional time-independent
Schr\"odinger equation \cite{stockmann,maryland,sridhar}.  Therefore,
by studying the modes of parallel-plate resonators we can gain insight
into the behavior of two-dimensional solutions to the Schr\"odinger
equation.

The resonator consisted of two parallel copper plates, 1 meter square,
separated by a distance of 1.25~cm.  One side of the resonator
consisted of a copper wall.  The other three sides were lined with a
11.5~cm thick layer of microwave absorber (C-RAM LF-79, Cuming
Microwave Corp.)\ designed to provide 20~dB of attenuation in the
reflected wave intensity in the range 0.6-40~GHz.  The absorber
prevented outgoing waves from returning to the resonator, thereby
simulating an open system in the directions away from the wall.  An
antenna was inserted normal to the plates, 2~mm from the wall, to
simulate the QPC.  The curved reflector was formed from a rectangular
aluminum rod bent into an arc with radius of curvature $R = 30.5\text
{ cm}$.  Various opening angles $\alpha$ were used: $115^\circ$,
$112^\circ$, $109^\circ$, and $106^\circ$.

\begin{figure}
    \centerline{\epsfig{figure=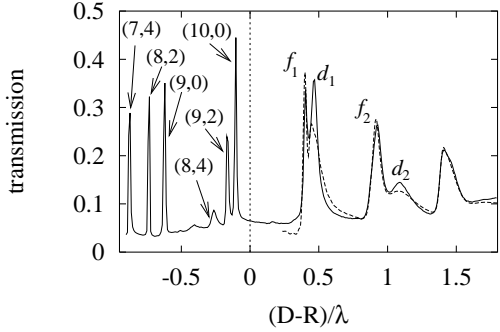,width=7cm}}
    \caption{
      Experimental transmission versus reflector-wall separation at a
      fixed frequency of 5.63 GHz; i.e., $R = 5.7 \lambda$.  The
      stable/unstable transition point occurs at abscissa zero.  In
      the stable regime, the peaks are labeled by their radial and
      angular quantum numbers $(n,m)$.  In the unstable regime, the
      diffractive resonances (labeled $d$) appear to the right of the
      Fabry-Perot peaks (labeled $f$). The dashed curve is the result
      of a semiclassical calculation which takes diffractive orbits
      into account (see text). The opening angle for the reflector was
      $\alpha = 106^\circ$.}
    \label{transmission}
\end{figure}

Instead of measuring the transmission of the resonator, we measured
the reflection back to the antenna; for this we used an HP8720D
network analyzer in `reflection' mode (the complex $S_{11}$ parameter
of the resonator was measured).  We inferred the transmission
probability $|T|^2$ via $|T|^2 = 1 - |R|^2$, where $R = S_{11}$ is the
measured reflection coefficient.  Because of the proximity of the
antenna to the wall, it was only weakly coupled to the resonator;
therefore, in the absence of the reflector, the transmission
coefficient was close to zero.  However, when the reflector was
present, the transmission experienced pronounced maxima at certain
frequencies.  In Fig.~\ref{transmission} we show a transmission
spectrum at fixed frequency, as the distance between the wall and
reflector is varied.
In the unstable regime, there are two types of resonance.  The first
type, labeled $f$ in Fig.~\ref{transmission}, is related to the
horizontal orbit along the axis of symmetry, and bears some
resemblance to a Fabry-Perot type resonance between two half-silvered
mirrors.  The second type, labeled $d$, is supported by diffraction
off the tips of the reflector.

The wavefunctions corresponding to peaks $f_1$ and $d_1$ were
measured using the technique of Maier and Slater~\cite{maier}.  They
showed that the frequency shift of a given resonance due to a small
sphere of radius $r_0$ at a position $(x,y)$ is given by
\begin{equation}
    \frac{\omega^2 - \omega_0^2}{\omega_0^2} = 
    4\pi r_0^3 \left(\frac{1}{2}H_0^2(x,y) - E_0^2(x,y) \right), 
    \label{eq:sphere-shift}
\end{equation}
where $E_0$ and $H_0$ are the unperturbed electric and magnetic
fields.  Thus, the frequency shift is proportional to the local
intensity of the microwave field, and by measuring the shift as a
function of the position of the sphere, the field intensity of a
particular mode can be mapped out.  Note that the frequency shift will
be positive in regions where the magnetic field is large, and negative
where the electric field is large.  Also, the factor of $1/2$
multiplying the magnetic field in Eq.~(\ref{eq:sphere-shift})
indicates that the sphere is a stronger perturbation to the electric
field that magnetic field.  In our measurements, we found this to be
the case: the shifts were predominantly negative.  Appreciable
positive shifts were only found at the nodes of the electric field,
corresponding to maxima of the magnetic field.

\begin{figure}
    \centerline{\epsfig{figure=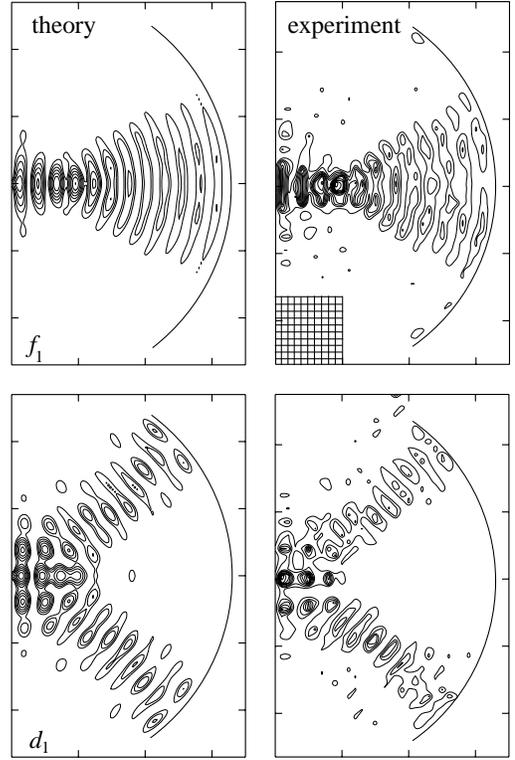,width=7cm}}
    \caption{
      Comparison between theoretical quantum wavefunctions (left) and
      experimentally measured microwave frequency shifts (right).  The
      two modes correspond to peaks $f_1$ and $d_1$, respectively, in
      Fig.~\ref{transmission}.  The wall is located on the left
      vertical axis in each plot, and the reflector position is
      indicated by the arc.  The graph ticks are 10 cm apart.  The
      fine grid indicates the spacing of the experimentally sampled
      points (grid spacing 1~cm).}
    \label{wavefunctions} 
\end{figure}

Figure~\ref{wavefunctions} shows theoretical quantum wavefunctions
compared with experimentally measured frequency shifts for the
resonances labeled by $f_1$ and $d_1$ in Fig.~\ref{transmission}.
The measured frequency shift is plotted as a function of sphere
position.  It is important to note that the frequency shift is not
proportional to $E^2$, but rather to $H^2/2 - E^2$.  Therefore we show
only negative contour lines below 20\% of the maximum negative shift,
and thereby emphasize regions of strong electric field.  The
similarity between theory and experiment is striking.

The wavefunction labeled $f_1$ in Fig.~\ref{wavefunctions} is clearly
associated with the horizontal orbit along the axis of symmetry.  Rays
emanating from a point source located on the axis of symmetry next to
the wall bounce off the reflector and come to an approximate focus
about 10~cm from the source.  The focus is approximate because of
spherical (or in this case cylindrical) aberration.

Now we turn our attention to the state labeled $d_1$ in
Fig.~\ref{wavefunctions}.  As noted above, the only periodic orbit in
the unstable regime is the horizontal orbit, along the axis of
symmetry.  The pictured wavefunction, however, clearly has very little
amplitude along this periodic orbit.  Instead the wavefunction has a
band of higher amplitude running from the region of the tip of the
mirror to the QPC, but in the unstable regime there is no {\em
  classical} periodic orbit that does this.  Theoretical studies have
suggested that states such as $d_1$ are supported by orbits that
undergo {\em diffraction} off the tips of the
reflector~\cite{jonathan}.  One such orbit is shown in
Fig.~\ref{schematic}(b).  Rays that hit the smooth surfaces of the
reflector or wall undergo specular reflection, whereas the rays that
hit near the reflector tips can be diffracted.  A fraction of the wave
amplitude can then return to the QPC from this region, thus setting up
a {\em non-classical\/} closed orbit.  All peaks labeled with a $d$ in
Fig.~\ref{transmission} are supported by such diffractive orbits.

Numerical calculations have shown that for energies off resonance, the
quantum wavefunction is often intermediate between those shown for
$f_1$ and $d_1$, in the sense that amplitude seems to be running from
the QPC to some point between the center of the mirror and the
tip~\cite{jonathan}.  This can be understood in terms of the
interference of paths with each other as they ``walk off'' the
horizontal orbit and escape the resonator.  Thus diffraction does not
play a major role in determining the off-resonance wavefunctions.
However, diffraction {\em is} instrumental in determining the
on-resonance wavefunctions underlying the conductance peaks $d_1$ and
$d_2$ in Fig.~\ref{transmission}.

\begin{figure}
    \centerline{\epsfig{figure=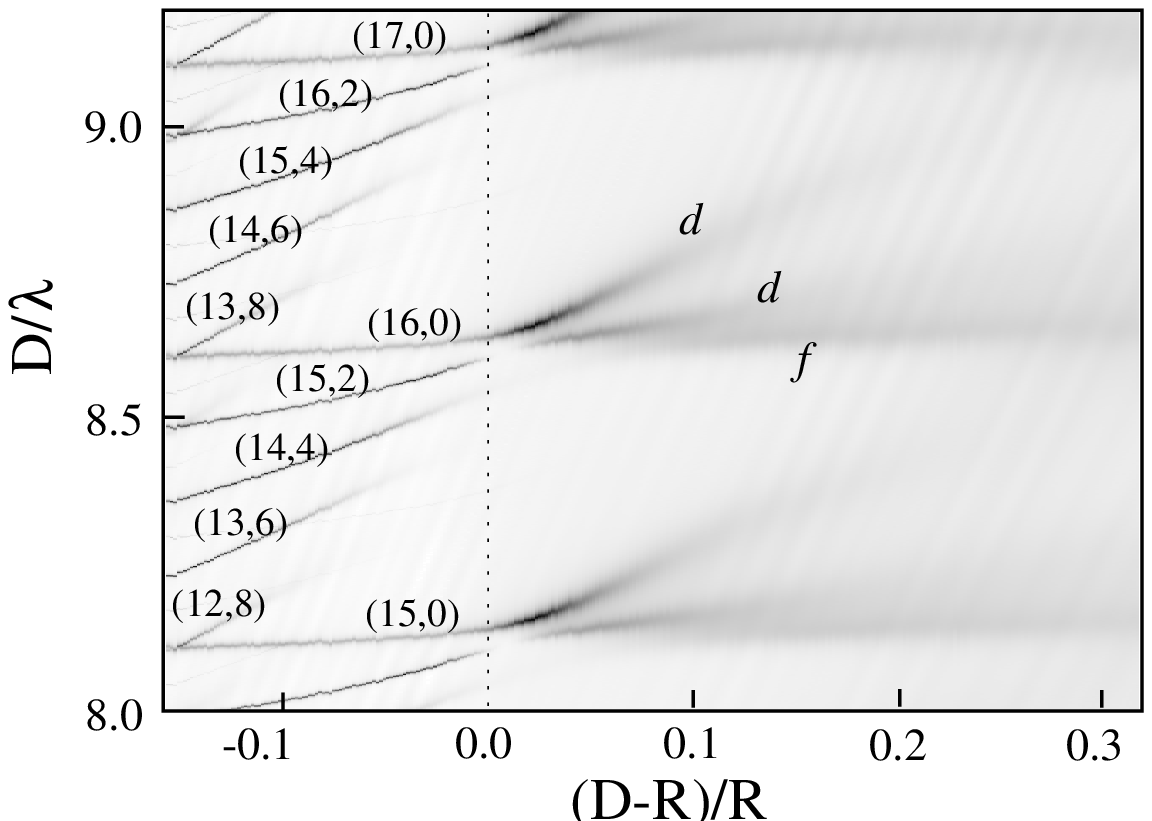,width=9cm}}
    \caption{
      Experimental transmission versus reflector-wall separation and
      wavelength.  High transmission regions are dark.  On the left of
      the vertical dotted line is the stable regime, where the
      transmission peaks are sharp.  The quantum numbers $(n,m)$ are
      indicated for a few peaks.  On the right is the unstable regime,
      where the resonances become wider and diffractive orbits become
      important.  Transmission peaks supported by diffractive orbits
      are marked by `d'.}
    \label{pretty}
\end{figure}

Figure~\ref{pretty} shows a more global picture of the transmission
properties of the resonator.
Here we plot the transmission of the resonator as both the wavelength
and the reflector-wall separation are varied.  Each vertical slice
through this figure is a frequency spectrum with fixed reflector
position; the dotted line marks the classical transition from stable
to unstable motion that occurs when the reflector's center of curvature
moves to the right of the QPC.  The vertical axis indicates how many
wavelengths fit along the horizontal orbit between the QPC and the
reflector.  The repetition of the resonance pattern every
half-wavelength in the vertical direction is analogous to the
half-wavelength periodicity of a Fabry-Perot cavity.

In the stable regime we have labeled the peaks with their quantum
numbers, $(n,m)$.  The vertical axis is chosen to make the $m = 0$
resonance peaks approximately horizontal in this figure.  As the
stable/unstable transition is approached, the peaks with high $m$
disappear one by one because their large angular sizes allow them to
escape around the reflector.

At the stable/unstable transition, all of the resonances in a family
would be approximately degenerate, but instead there is an avoided
crossing.  The level repulsion is caused by a coupling that is partly
mediated by diffraction; this subject will be explored more thoroughly
in a future publication.

In the unstable regime, the only remaining classical periodic orbit is
the horizontal orbit, which itself becomes unstable.  The Fabry-Perot
peak (labeled $f$) is essentially quantized along the horizontal
orbit, so its position shows a simple dependence on reflector
position.  It becomes broad in the unstable regime, with a lifetime
given by the classical Lyapunov stability exponent of the horizontal
orbit.  Two diffractive resonances (labeled by $d$), are also visible;
they separate from the Fabry-Perot type peak as the reflector is moved
away from the wall.  If the angle $\alpha$ subtended by the mirror is
changed, the position of the Fabry-Perot peak remains unaffected
whereas the diffractive peaks shift.

The diffractive peaks labeled by $d$ in Fig.~\ref{pretty} cannot be
explained by semiclassical theory unless diffraction off the tips of
the reflector is included.  The semiclassical calculation involves
launching a manifold of rays from the QPC, tracking their phases as
they reflect off the reflector and cross caustics or foci, and then
adding coherently the amplitudes of any orbits that return to the QPC.
To include diffraction, we also allow for the fact that every ray that
hits the tip of the reflector is scattered in all directions, with an
angle-dependent amplitude~\cite{SC-diffraction}.  Any of those
scattered rays that return to the QPC gives an additional contribution
to the conductance.  The details of the semiclassical theory including
diffraction will be presented in a future paper.

\begin{figure}
    \centerline{\epsfig{figure=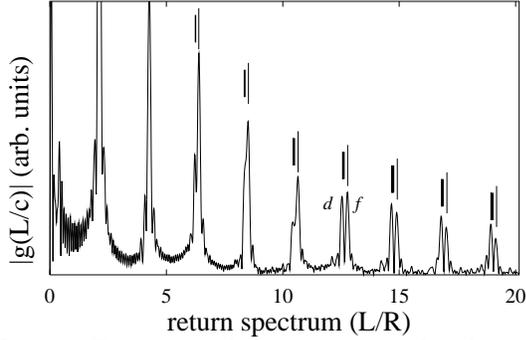,width=7cm}}
    \caption{
      Experimental return spectra for the unstable regime. Time has
      been converted to the ratio $L/R$, where $L$ is the length of
      the orbit.  The splitting of the peaks clearly demonstrates the
      influence of both Fabry-Perot type orbits (marked `$f$') and
      diffractive orbits (marked `$d$'), which are slightly shorter.
      The calculated lengths of the orbits are shown by vertical bars;
      short bars for the diffractive orbits, and longer bars for the
      horizontal orbit.  For these plots the opening angle was
      $115^\circ$, and the reflector-wall separation was 32.5~cm.}
    \label{fourier}
\end{figure}

Further evidence of the importance of diffractive orbits is contained
in the return spectrum (Fig.~\ref{fourier}), which is the Fourier
transform of the complex reflection scattering matrix element
$S_{11}(\omega)$.  That is, if a short pulse were emitted from the
antenna at time $t = 0$, echos would return to the antenna at certain
later times.  These echos are indicated by peaks in the return
spectrum.  Many of the return peaks are split due to the coexistence
of the horizontal orbit and diffractive orbits with slightly shorter
return times.  The horizontal orbit and its repetitions, which have
lengths indicated in Fig.~\ref{fourier} by long vertical bars, cause
the primary peaks in each group.  In addition, near each primary peak
there is a family of diffractive orbits (with lengths indicated in
Fig.~\ref{fourier} by short vertical bars) which combine to form a
secondary peak.  The presence of this splitting in the return spectrum
is strong evidence in support of the claim that diffraction off the
edges of the reflector supports other closed orbits, which lead to
resonances in the transmission spectra.  Note that for the long
orbits, the diffractive peaks are even stronger than the peaks from
the geometric orbit. This is because the number of diffractive orbits
increases linearly with the length of the orbit, whereas there is
alwayse only one geometric orbit, regardless of length.

In summary, we have demonstrated the existence of diffractive orbits
in an open microwave billiard, which give rise to wavefunctions that
would not be predicted by a simple semiclassical theory.  Such orbits
are of importance in open, unstable systems where the number of
unstable classical periodic orbits is small.  In such systems,
diffraction can play a major role in determining the spectrum of the
system.

We thank the Hewlett Packard Corporation for the loan of a network
analyzer that was used in these experiments.  We thank J.~D. Edwards
for the computer program that was used for the quantum computations.
This work was supported through funding from Harvard University,
ITAMP, and also Grant No.\ NSF-CHE9610501.


\end{document}